\documentclass[11pt]{article}
\usepackage{amsmath,amssymb,color,epsfig,cite}
\usepackage{graphicx}
\usepackage{subfigure}
\usepackage{setspace}

\usepackage{amsfonts}

\newcommand{\hoch}[1]{$\, ^{#1}$}

\makeatletter
\@addtoreset{equation}{section}
\makeatother

\newcommand{\be}{\begin{equation}}
\newcommand{\ee}{\end{equation}}
\newcommand{\bea}{\setlength\arraycolsep{2pt} \begin{eqnarray}}
\newcommand{\eea}{\end{eqnarray}}
\newcommand{\nn}{\nonumber}

\def\ft#1#2{{\textstyle{\frac{\scriptstyle #1}{\scriptstyle #2} } }}
\def\fft#1#2{{\frac{#1}{#2}}}

\def\0{{\sst{(0)}}}
\def\1{{\sst{(1)}}}
\def\2{{\sst{(2)}}}
\def\3{{\sst{(3)}}}
\def\4{{\sst{(4)}}}
\def\5{{\sst{(5)}}}
\def\6{{\sst{(6)}}}
\def\7{{\sst{(7)}}}
\def\8{{\sst{(8)}}}
\def\9{{\sst{(9)}}}

\def\sst#1{{\scriptscriptstyle #1}}

\begin{document}



\begin{center}
{\large {\bf Universal Structure of Covariant Holographic Two-Point Functions\\
In Massless Higher-Order Gravities}}

\vspace{10pt}
Yue-Zhou Li\hoch{\dag},  H. L\"u\hoch{*},  Zhan-Feng Mai\hoch{\ddag}

\vspace{15pt}

{\it Center for Joint Quantum Studies and Department of Physics,\\
School of Science, Tianjin University, Tianjin 300350, China}

\vspace{30pt}

\underline{ABSTRACT}
\end{center}

We consider massless higher-order gravities in general $D=d+1$ dimensions, which are Einstein gravity extended with higher-order curvature invariants in such a way that the linearized spectrum around the AdS vacua involves only the massless graviton.  We derive the covariant holographic two-point functions and find that they have a universal structure.  In particular, the theory-dependent overall coefficient factor $\mathcal{C}_T$ can be universally expressed by $(d-1) \mathcal{C}_T=\ell (\partial a/\partial\ell)$, where $a$ is the holographic $a$-charge and $\ell$ is the AdS radius. We verify this relation in quasi-topological Ricci polynomial, Einstein-Gauss-Bonnet, Einstein-Lovelock and Einstein cubic gravities.  In $d=4$, we also find an intriguing relation between the holographic $c$ and $a$ charges, namely $c=\ft13\ell (\partial a/\partial \ell)$, which also implies $\mathcal{C}_T=c$.

\vfill {\footnotesize \hoch{\dag}liyuezhou@tju.edu.cn\ \ \ \hoch{*}mrhonglu@gmail.com\ \ \
\hoch{\ddag}makey@mail.bnu.edu.cn}

\pagebreak

\tableofcontents
\addtocontents{toc}{\protect\setcounter{tocdepth}{2}}


\newpage

\section{Introduction}
Since it was proposed decades ago, the AdS/CFT correspondence \cite{Maldacena:1997re} has been widely employed to deal with strongly coupled conformal field theories (CFT) by studying their dual classical anti-de Sitter (AdS) gravities. For instance, arising from the holographic renormalization procedure \cite{deBoer:2000cz,Bianchi:2001kw,Skenderis:2002wp}, the AdS/CFT correspondence provides a more manageable approach to compute $n$-point functions \cite{Gubser:1998bc,Witten:1998qj,Freedman:1998tz,Liu:1998bu,Muck:1998rr,KeskiVakkuri:1998nw}. The basic idea of the AdS/CFT correspondence is the field-operator duality \cite{Witten:1998qj}, which states that any field $\phi$ defined in the bulk has the corresponding gauge invariant operator $\mathcal{O}_{\phi}$ defined in the boundary. The boundary value $\phi_0$ of $\phi$ is identified as the source coupled to the operator, and furthermore the partition function of the boundary CFT$_d$ is identified with the on-shell action of AdS$_{d+1}$ gravities \cite{Gubser:1998bc,Witten:1998qj}
\be
S_{\rm gr}\Big|_{\phi_0}=\langle \exp(-\int d^dx\, \phi_0 \mathcal{O}_{\phi})\rangle\,.\label{identi}
\ee
With this identification, the $n$-point functions of the operator $\mathcal{O}_\phi$ of the CFT$_d$ can be computed by evaluating the on-shell action of the gravity theories:
\be
\langle\mathcal{O}_{\phi}(x_1)\cdots\mathcal{O}_{\phi}(x_n)\rangle\sim\fft{\delta^n S}{\delta {\phi_0}(x_1)\cdots\delta{\phi_0}(x_n)}\sim \fft{\delta^{n-1} \langle\mathcal{O}_\phi(x_1)\rangle}{\delta {\phi_0}(x_2)\cdots\delta{\phi_0}(x_n)} \,.\label{dic}
\ee
In pure gravities, the only available field is the metric, and the corresponding operator is the energy-momentum tensor of the boundary CFT. The corresponding two-point functions in Einstein gravities in general dimensions were previously obtained \cite{Liu:1998bu}.

In this paper, we consider Einstein gravity in $D=d+1$ dimensions, extended with higher-order curvature invariants.  Quadratically-extended gravity was shown to be renormalizable in four dimensions, but the theory contains additional massive scalar and ghostlike massive spin-2 modes \cite{Stelle:1976gc}. There exists a critical point in the parameter space such that the massive spin-2 modes become log modes \cite{Lu:2011zk,Deser:2011xc}.  The covariant two-point functions in four-dimensional critical gravity were obtained in \cite{Johansson:2012fs}. Analogous two-point functions in four-dimensional conformal gravity were also obtained in \cite{Ghodsi:2014hua}.

In this paper, we shall focus on higher-order gravities whose linearized spectrum around the AdS vacuum contains only the (massless) graviton.  In other words, we consider special classes of the higher-order curvature invariants such that the massive scalar and spin-2 modes are decoupled from the linearized theory around the AdS vacuum.  Such combinations include the well-known Gauss-Bonnet and the more general Lovelock series \cite{ll}.  In fact, if we consider only Riemann curvature polynomials, then the linearized spectrum on the AdS vacuum contains at most an additional massive scalar and/or a massive spin-2 mode, the decoupling of these two modes requires only two linear constraints on the coupling constants. (A comprehensive discussion in this context can be found in \cite{Tekin:2016vli,Bueno:2016xff,Bueno:2016ypa}.) For the quadratic Riemann invariants, these two constraints lead to the Gauss-Bonnet combination. For cubic Riemann invariants, which have a total of eight independent structures, the constraints lead to a six-parameter family of combinations, including the cubic Lovelock terms.
If one considers only Ricci polynomials, the resulting theory is quasi-topological in that the linearized theory around the AdS vacuum is identical to that of pure Einstein gravity \cite{Li:2017ncu}.
We refer all these theories to ``massless higher-order gravities.'' The subject has attracted considerable attention and many massless higher-order gravities were constructed in literature \cite{Dehghani:2011vu,Karasu:2016ifk,Bueno:2016dol,Cisterna:2017umf,Hennigar:2017ego,Ahmed:2017jod}.

In this paper, we compute covariant holographic two-point functions in massless higher-order gravities. We follow the method of \cite{Johansson:2012fs,Ghodsi:2014hua}, developed for four dimensions.  We generalize to general $D=d+1$ dimensions and obtain explicit covariant two-point functions for the boundary energy-momentum tensor.  The results can be written compactly as
\bea
&&\langle T_{ij}(x)T_{kl}(0)\rangle =\fft{N_2\,\mathcal{C}_T\,
\mathcal{I}_{ijkl}(x)}{x^{2d}}\,,
\label{2-pt-uni}
\eea
where $N_2$, given by section \ref{two-point-result}, is a numerical constant depending only on $d$. The boundary spacetime tensor $\mathcal{I}_{ijkl}(x)$ is defined by
\bea
\mathcal{I}_{ijkl}(x)=\ft{1}{2}\big(I_{ik}(x)I_{jl}(x)+I_{il}(x)I_{jk}(x)\big)-
\ft{1}{d}\eta_{ij}\eta_{kl}\,,\qquad I_{ij}(x)=\eta_{ij}-\fft{2x_{i}x_{j}}{x^2}\,,\label{Iijkl}
\eea
where $x_i$'s are the cartesian coordinates of the boundary Minkowski spacetime $\eta_{ij}$. The structure matches the results of CFT \cite{Osborn:1993cr,Erdmenger:1996yc,Coriano:2012wp}, and also matches the results of Einstein gravity \cite{Liu:1998bu} and Gauss-Bonnet gravity \cite{Buchel:2009sk}. In literature, the overall coefficient $N_2 {\cal C}_T$ is denoted as $C_T$.  In this paper, we strip off the inessential (in the current context,) universal numerical constant $N_2$ from $C_T$ and focus on the dimensionful quantity ${\cal C}_T$ instead.

The coefficient $\mathcal{C}_T$ depends on the detail of the theory; however, for massless higher-order gravities, we find that there is a universal expression relating $\mathcal{C}_T$ to the $a$-charge:
\be
\mathcal{C}_T=\fft{1}{d-1}\,\ell \fft{\partial a}{\partial \ell}\,,\label{CT-uni}
\ee
where $a$ is the coefficient of the Euler density in the holographic conformal anomaly, and is typically referred to as the $a$-charge. (See, e.g. \cite{Henningson:1998gx,Henningson:1998ey,Imbimbo:1999bj,Nojiri:1999mh,Blau:1999vz}.) The parameter $\ell$ is the radius of the AdS vacuum. It is important to emphasize that the $a$-charge in (\ref{CT-uni}) must be expressed in terms of $\ell$ and the bare coupling constants of the higher-order curvature invariants as independent parameters, with the bare cosmological constant $\Lambda_0$ solved in terms of these quantities by the equations of motion,
as was done in \cite{Li:2017txk}.  This way of expressing the $a$-charge is different from those in literature (e.g.~\cite{Buchel:2009sk}) where the bare cosmological constant is typically an explicit parameter in the expression.  It is also worth pointing out that the inessential and universal numerical dependence on $\pi$ and $d$ were stripped off from our definitions of ${\cal C}_T$ and the $a$-charge so that the relation (\ref{CT-uni}) is simple.

Even though the concept of anomaly makes sense only in odd bulk $D$ dimensions, the central charge $a$ can be nevertheless defined in general dimensions \cite{Li:2017txk,Myers:2010xs,Myers:2010tj}. We verify the expression (\ref{CT-uni}) for Einstein gravity, quasi-topological Ricci polynomial gravities \cite{Li:2017ncu}, Einstein-Gauss-Bonnet gravity, and more general Lovelock gravities \cite{ll} and furthermore Einstein-cubic gravities \cite{Li:2017txk}.

The paper is organized as follows. In section \ref{cov-stru}, we solve the full linear massless graviton perturbations around the AdS vacuum of flat Minkowski boundary in general dimensions. We give a well-defined basis to express the solutions covariantly and compactly. Under the basis we adopt, we obtain the covariant and universal structure of two-point functions in general dimensions. We find Einstein gravity and general quasi-topological Ricci polynomial gravities satisfy the relation (\ref{CT-uni}) directly. In section \ref{inclu-GB}, we study Einstein gravity extended with the Gauss-Bonnet term. We perform the Fefferman-Graham (FG) expansion of the flat AdS boundary to obtain the one-point function, and read off $\mathcal{C}_T$ in two-point functions. We demonstrate that (\ref{CT-uni}) is valid. In section \ref{inclu-ll}, we consider general Einstein-Lovelock gravities. We obtain $\mathcal{C}_T$, calculate the holographic $a$-charge and show that (\ref{CT-uni}) is indeed satisfied. In section \ref{cubicgr}, we repeat the verification for the general Einstein-Riemann cubic gravities where the massive modes are decoupled.  We conclude the paper in section \ref{conc}. In appendix \ref{B2}, we exhibit the details of the asymptotic behavior of the linear perturbative solutions.

\section{The Covariant Structure of Two-Point Functions}
\label{cov-stru}

\subsection{The holographic dictionary}

We begin with a brief review of how to compute the two-point function based on the holographic dictionary. We assume that our gravity theory admits an AdS vacuum with radius $\ell$.  Furthermore, we restrict ourselves to consider the AdS spacetime with the flat boundary throughout this paper. The metrics of the asymptotic AdS in $D=d+1$ dimensions take the form
\be
ds^2=\fft{\ell^2}{r^2}dr^2+h_{ij}dx^i dx^j=\fft{\ell^2}{r^2}dr^2+r^2 g_{ij}dx^i dx^j\,.\label{FG-co}
\ee
In this coordinate system, the AdS boundary is located at $r\rightarrow \infty$. At the asymptotic region, the FG expansion of $g_{ij}$ is
\be
g_{ij}=g^{(0)}_{ij}+\fft{g^{(d)}_{ij}}{r^d}+\cdots\,,\label{FGflat}
\ee
for theories with only the massless graviton modes. The leading $g^{(0)}_{ij}$ is interpreted as the source of the boundary CFT in the context of the holographic dictionary \cite{Skenderis:2002wp}, and the two-point function can be obtained, as by (\ref{dic}),
\be
\langle T_{ij}T_{kl}\rangle=2\fft{\delta\langle T_{ij}\rangle}{\delta g^{(0)kl}}\,.\label{2-pt-0}
\ee
Therefore, for the purpose of computing the two-point functions, we shall first obtain the one-point function of the holographic energy-momentum tensor. It turns out to be \cite{deHaro:2000vlm}
\be
\langle T_{ij}\rangle=T_{ij}(h)r^{d-2}\big|_{r\rightarrow\infty}=-\fft{2}{\sqrt{g^{(0)}}}\fft{\delta S}{\delta g^{(0)ij}}\,,\label{1-pt-0}
\ee
where $T_{ij}(h)$ is the Brown-York energy-momentum tensor associated with the embedding boundary metric $h_{ij}$
\be
T_{ij}(h)=-\fft{2}{\sqrt{h}}\fft{\delta S}{\delta h^{ij}}\,.\label{BY}
\ee
The holographic dictionary \cite{Skenderis:2002wp} then gives us
\be
\langle T_{ij} \rangle\sim g^{(d)}_{ij}\,.
\ee
For example, the one-point function of pure Einstein gravity is given by \cite{deHaro:2000vlm,Fischetti:2012rd}
\be
\langle T_{ij}\rangle=\fft{d}{16\pi\ell}g^{(d)}_{ij}\,.\label{1-pt-Ein}
\ee
Thus the computation of the two-point function (\ref{2-pt-0}) now involves the evaluation of the quantity
\be
\fft{\delta g^{(d)}_{ij}}{\delta g^{(0)kl}}\,.\label{inv-cal}
\ee
In other words, the main task is to determine how the response mode $g^{(d)}_{ij}$ depends on the source $g^{(0)}_{ij}$ around the AdS vacuum. In order to obtain the result covariantly, we closely follow the method developed in \cite{Johansson:2012fs,Ghodsi:2014hua} for $D=4$ dimensions.  We generalize the method to general $D=d+1$ dimensions and obtain the complete linear perturbations around the AdS backgrounds, and then rewrite solutions in terms of the metric basis covariantly and compactly. This enables us to read off (\ref{inv-cal}) explicitly.

\subsection{Linear perturbations and graviton modes}
\subsubsection{Transverse-traceless modes}

In order to evaluate (\ref{inv-cal}), we start with the pure AdS background with the Minkowski boundary, i.e. $h_{ij}=r^2 \eta_{ij}$ in (\ref{FG-co}). We turn on the perturbation $\hat{h}_{ij}=r^2 f_{ij}$ so that the metric is
\be
ds^2=\fft{\ell^2}{r^2}dr^2+r^2\eta_{ij}dx^i dx^j+\hat{h}_{ij}dx^i dx^j\,.
\ee
In this paper, we consider Einstein gravity extended with higher-order curvature invariants.  The linear spectrum in general contains massive scalar and spin-2 modes, in addition to the usual massless graviton.  We shall restrict ourselves to consider theories where these additional modes are decoupled and only the graviton survives. These theories were referred to as massless higher-order gravities in the introduction. The decoupling of the ghostlike massive spin-2 modes is clearly necessary for a healthy theory; it was shown that it is also necessary for the massive scalar modes to be decoupled in order for the theory to have a holographic $a$-theorem \cite{Li:2017txk}.  For massless higher-order gravities, the linearized equations of motion around the AdS vacuum is simply of the graviton mode. For simplicity, we restrict ourselves to the transverse-traceless gauge, in which the perturbation satisfies the conditions
\be
\nabla^{j}\hat{h}_{ij}=0\,,\qquad \hat{h}=0\,.\label{gauge}
\ee
The linearized equation is then given by
\be
\kappa_{\rm eff}(\widetilde{\Box}+\fft{2}{\ell^2})\hat{h}_{ij}=0\,,\label{linear}
\ee
where $\kappa_{\rm eff}$ is given by
\be
\kappa_{\rm eff} = \fft{1}{16\pi} + \cdots\,.
\ee
The first term comes from Einstein gravity where the Newton's constant is set to unity, and the ellipses denote the the contributions from the higher-order curvature invariants.  The Laplacian $\widetilde{\Box}$ in (\ref{linear}) is defined with respect to the pure AdS background. Explicitly, we have
\be
\fft{r^2}{\ell^2}\hat{h}''_{ij}
+(d-3)\fft{r}{\ell^2}\hat{h}'_{ij}-2(d-2)\fft{1}{\ell^2}\hat{h}_{ij}
+\fft{1}{r^2}\Box\hat{h}_{ij}=0\,,
\ee
where $\Box$ is the Laplacian with respect to the flat Minkowski metric $\eta_{ij}$. The equation can be solved completely by separation of variables
\be
\hat{h}_{ij}=e^{-i p\cdot x}\Psi_{ij}(r)\,.
\ee
For convenience, we first consider the time-like modes and we choose $p_i=E\delta_i^t$.  The solutions are
\be
f_{ij}=e^{-i E t}\,r^{-\fft{d}{2}}\Big(c^{a}_{ij}J_{\fft{d}{2}}(\fft{E\ell}{r})+
c^b_{ij}Y_{\fft{d}{2}}(\fft{E\ell}{r})\Big)\,,
\label{pertur-solu1}
\ee
where $J$ and $Y$ are the first and second kind Bessel functions respectively and ($c_{ij}^a$, $c_{ij}^b$) are integration constants. Thus we see that the perturbative functions $f_{ij}$ can be expanded as
\be
f_{ij}=f^{(0)}_{ij}+\cdots+\fft{f^{(d)}_{ij}}{r^d}+\cdots\,.\label{f-FG}
\ee
In this context, we have
\be
\delta g^{(0)}_{ij}=f^{(0)}_{ij}\,,\qquad \delta g^{(d)}_{ij}=f^{(d)}_{ij}\,.\label{identify}
\ee
Note that the terms in the first ellipses of (\ref{f-FG}), which are absent in (\ref{FGflat}), are due to the fact that the boundary metric becomes no longer flat with the perturbations.

The quantity (\ref{inv-cal}) is not yet well defined for general integration constants $c_{ij}^a$ and $c_{ij}^b$.  One needs to impose further an in-falling boundary condition at the Poincare horizon $r\rightarrow 0$, namely
\be
f_{ij}\Big|_{r\rightarrow 0}\sim e^{-{\rm i} E(t-\fft{\ell}{r})}\,.
\ee
Approaching to $r\rightarrow0$, the Bessel functions $J$ and $Y$ have the following asymptotic behaviors
\be
J_{\fft{d}{2}}(\fft{E\ell}{r})\sim\sqrt{\fft{2r}{\pi\ell E}}\cos\big(\fft{E\ell}{r}-\fft{d\pi}{4}-\fft{\pi}{4}\big)
\,,\qquad Y_{\fft{d}{2}}(\fft{E\ell}{r})\sim\sqrt{\fft{2r}{\pi\ell E}}\sin\big(\fft{E\ell}{r}-\fft{d\pi}{4}-\fft{\pi}{4}\big)\,,
\ee
we therefore must impose
\be
c^{b}_{ij}={\rm i} c^{a}_{ij}\,.\label{in-falling}
\ee
The solutions with the right boundary and gauge conditions (\ref{gauge}) are
\be
f_{ij}=e^{-{\rm i} E t}\,r^{-\fft{d}{2}}c^{a}_{ij}H_{\fft{d}{2}}(\fft{E\ell}{r})\,,\qquad c^a_{it}=0\,,\qquad
c_{ij}^a\eta^{ij}=0\,,
\label{pertur-solu2}
\ee
where $H$ is the first kind Hankel function. It is clear that we have a total of $(d-2)(d+1)/2$ independent physical modes under the transverse-traceless gauge.

\subsubsection{PBH transformation - all modes}

We obtained the $(d-2)(d+1)/2$ solutions of the linearized Einstein equation (\ref{linear}) in the transverse-traceless gauge. It is necessary to find out the full $d(d+1)/2$ modes in order to obtain the covariant two-point functions. The missing modes are pure gauge and can be generated by the PBH transformation, which refers to a particular diffeomorphism that remains within the FG coordinates \cite{Imbimbo:1999bj}
\be
\hat{h}_{\mu\nu}\rightarrow  \hat{h}_{\mu\nu}+\nabla_{\mu}\xi_{\nu}+\nabla_{\nu}\xi_{\mu}\,,\qquad
\hat{h}_{rr}=0\,,\qquad \hat{h}_{ir}=0\,.\label{PBH}
\ee
The ansatz for PBH transformation is given by
\be
\xi_{\mu}=e^{-{\rm i} E t}\hat{\xi}_{\mu}\,.
\ee
To avoid ambiguity, we denote $\tilde{i}$ as the space-like indices on the boundary, i.e~$i=(t,\tilde{i})$. The PBH equations in (\ref{PBH}) thus imply
\bea
\hat{h}_{rr}=0:&&\qquad\hat{\xi}'_r+\fft{1}{r}\hat{\xi}_r=0\,,\cr
\hat{h}_{\tilde{i}r}=0:&&\qquad\hat{\xi}'_{\tilde{i}}-\fft{2}{r}\hat{\xi}_{\tilde{i}}=0\,,\cr
\hat{h}_{tr}=0:&&\qquad  -({\rm i} E\,\hat{\xi}_r+\fft{2}{r}\hat{\xi}_t)+\hat{\xi}'_t=0\,.
\eea
These equations can be solved explicitly:
\be
\hat{\xi}_r=\fft{b_r}{r}\,,\qquad \hat{\xi}_{\tilde{i}}=b_{\tilde{i}} r^2\,,\qquad \hat{\xi}_t=b_t r^2-\fft{1}{2}b_r E\,,\label{PBH-solu}
\ee
where $(b_r,b_t,b_{\tilde i})$ are the integration constants associated with the $d+1$ pure gauge modes.
Substituting (\ref{PBH-solu}) into the first equation of (\ref{PBH}), and recalling the original transverse and traceless solutions (\ref{pertur-solu2}), we arrive at the full set of the $d(d+1)/2$ solutions
\bea
&& f_{tt}=-e^{-{\rm i} E t}\Big(2{\rm i} b_t E+b_r(\fft{2}{\ell^2}+\fft{E^2}{r^2})\Big)\,,\qquad f_{t \tilde{i}}=-{\rm i} e^{-{\rm i} E t}b_{\tilde{i}} E\,,\nn\\
&&
f_{\tilde{i}\tilde{j}}=e^{-{\rm i} E t}\Big(r^{-\fft{d}{2}}c^a_{\tilde{i}\tilde{j}}H_{\fft{d}{2}}(\fft{E\ell}{r})+\fft{2}{\ell^2}b_r \delta_{\tilde{i} \tilde{j}}\Big)\,.\label{full-solu}
\eea

\subsubsection{Metric basis}

Although the integration constants $c_{\tilde i\tilde j}^a$, $b_t$, $b_{\tilde i}$ in the
solutions given by (\ref{full-solu}) carry the boundary spacetime indices, they are not the spacetime tensors.  It is thus necessary to construct a symmetric tensor basis in terms of which the solutions can be expressed.  For this purpose, we consider the vielbein $e^{(k)}_i$ for the Minkowski metric:
\be
\eta_{ij}=-e^{(0)}_i e^{(0)}_j+e^{(1)}_i e^{(1)}_j+\cdots+e^{(k)}_ie^{(k)}_j+\cdots +e^{(d-1)}_i e^{(d-1)}_j\,.
\ee
From these vielbein, we can construct the $d(d+1)/2$ symmetric tensors:
\bea
&& \mathcal{P}^{(k)}_{ij}=e^{(0)}_i e^{(k)}_j+e^{(k)}_i e^{(0)}_j\,,\qquad\qquad 1\leq k\leq d-1\,,
\cr && \mathcal{Q}^{(k)}_{ij}=e^{(1)}_i e^{(1)}_j-e^{(k+1)}_i e^{(k+1)}_j\,,\qquad 1\leq k\leq d-2\,,
\cr && \mathcal{L}^{(m,n)}_{ij}=e^{(m)}_i e^{(n)}_j+e^{(n)}_j e^{(m)}_i\,,~\qquad m\neq n~~~{\rm and}~~~1\leq m,n \leq d-1\,,
\cr && \mathcal{M}_{ij}=e^{(0)}_i e^{(0)}_j\,.\label{basis-1}
\eea
The full solutions in (\ref{full-solu}) can now be expressed as a matrix representation in terms of these boundary spacetime tensors:
\bea
&& f^{{\rm R}(k)}_{ij}=e^{-{\rm i} E t}E\,\mathcal{P}^{(k)}_{ij}\,,\qquad 1\leq k\leq d-1\,,\nn\\
&& f^{{\rm R}(k+d-1)}_{ij}=e^{-{\rm i} E t}r^{-\fft{d}{2}}H_{\fft{d}{2}}\big(\fft{E\ell}{r}\big)\mathcal{Q}^{(k)}_{ij}\,,\qquad 1\leq k\leq d-2\,,
\nn\\
&& f^{{\rm R}(k+2d-3)}_{ij}=e^{-{\rm i} E t}r^{-\fft{d}{2}}H_{\fft{d}{2}}\big(\fft{E\ell}{r}\big)\mathcal{L}^{(m,n)}_{ij}\,,\qquad 1\leq k\leq \ft12(d-1)(d-2)\,,\nn\\
&& f^{{\rm R}(\fft{1}{2}(d^2+d-2))}_{ij}=e^{-{\rm i} E t}E^2\mathcal{M}_{ij}\,,\qquad
f^{{\rm R}(\fft{1}{2}d(d+1))}_{ij}=e^{-{\rm i} E t}\Big(\fft{\eta_{ij}}{\ell^2}-\fft{E^2}{2r^2}\mathcal{M}_{ij}\Big)\,,\label{rep-1}
\eea
where the superscripts ${\rm R}(n)$, with $n=1,2,\cdots, d(d+1)/2$, denote the $n$'th representation. Note we have $p_i=E \delta_i^t$, this fact can be used to make (\ref{rep-1}) covariant, i.e.
\bea
&& f^{{\rm R}(k)}_{ij}=e^{-{\rm i} p\cdot x}\big(p_i e^{(k)}_j+p_j e^{(k)}_i\big)\,,
\quad\qquad\qquad\quad 1\leq k\leq d-1\,,\nn\\
&& f^{{\rm R}(k+d-1)}_{ij}=e^{-{\rm i} p\cdot x}r^{-\fft{d}{2}}H_{\fft{d}{2}}\big(\fft{|p|\ell}{r}\big)\mathcal{Q}^{(k)}_{ij}\,,\quad\qquad~ 1\leq k\leq d-2\,,\nn\\
&& f^{{\rm R}(k+2d-3)}_{ij}=e^{-{\rm i} p\cdot x}r^{-\fft{d}{2}}H_{\fft{d}{2}}\big(\fft{|p|\ell}{r}\big)\mathcal{L}^{(m,n)}_{ij}\,,\qquad 1\leq k\leq \ft12(d-1)(d-2)\,,\nn\\
&& f^{{\rm R}(\fft{1}{2}(d^2+d-2))}_{ij}=e^{-{\rm i} p\cdot x}p_i p_j\,,\qquad f^{{\rm R}(\fft{1}{2}d(d+1))}_{ij}=e^{-{\rm i} p\cdot x}(\fft{\eta_{ij}}{\ell^2}-\fft{1}{2r^2}p_i p_j)\,.\label{solufrep}
\eea
We now introduce the metric basis $E^I$ where the index $I$ runs from $1$ to $d(d+1)/2$
\bea
&& E^{k}_{ij}=p_i e^{(k)}_j+p_j e^{(k)}_i\,,\qquad 1\leq k\leq d-1\,,
\cr && E^{k+d-1}_{ij}=\mathcal{Q}^{(k)}_{ij}\,,\qquad\qquad~ 1\leq k\leq d-2\,,
\cr && E^{k+2d-3}_{ij}=\mathcal{L}^{(m,n)}_{ij}\,,\qquad\quad 1\leq k\leq \ft12(d-1)(d-2)\,,
\cr && E^{\fft{1}{2}(d^2+d-2)}_{ij}=p_i p_j\,,\qquad E^{\fft{1}{2}d(d+1)}_{ij}=\eta_{ij}\,.
\eea
However, this set of basis suffers from the fact that the traceless part is not orthogonal. To rescue this representation, we follow the Schmidt orthogonalization procedure to define a new basis, which is given by
\bea
&& \tilde{E}^{I}_{ij}=\sqrt{\fft{2(I-d+1)}{I-d+2}}\left(E^I_{ij}-\fft{1}{I-d+1}\sum_{K=d}^{I-1}E^K_{ij}\right)
\,,\qquad d\leq I\leq 2d-3\,,
\cr &&
\cr && \tilde{E}^I_{ij}=E^I_{ij}\,,\qquad {\rm other}\,\,\,I\,.\label{basis-2}
\eea
This set of basis satisfies the following important orthogonal condition which is useful for computing the two-point functions
\be
\tilde{E}^{I}_{ij}\tilde{E}_{K}^{ij}=2\delta^I_K\,,\qquad d\leq I\leq \fft{d^2+d-4}{2}\,,\qquad1\leq K \leq \fft{d(d+1)}{2}\,.\label{orth}
\ee
With this set of basis in hand, we can rewrite (\ref{solufrep}) as
\bea
&& f^{{\rm R}(k)}_{ij}=e^{-{\rm i} p\cdot x}\tilde{E}^k_{ij}\,,\qquad 1\leq k\leq d-1\,,\nn\\
&& f^{{\rm R}(k+d-1)}_{ij}=e^{-{\rm i} p\cdot x}r^{-\fft{d}{2}}H_{\fft{d}{2}}\big(\fft{|p|\ell}{r}\big)\tilde{E}^{k+d-1}_{ij}\,,
\qquad 1\leq k\leq d-2\,,\nn\\
&& f^{{\rm R}(k+2d-3)}_{ij}=e^{-{\rm i} p\cdot x}r^{-\fft{d}{2}}H_{\fft{d}{2}}\big(\fft{|p|\ell}{r}\big)\tilde{E}^{k+2d-3}_{ij}\,,\qquad 1\leq k\leq \ft12(d-1)(d-2)\,,\nn\\
&& f^{{\rm R}(\fft{1}{2}(d^2+d-2))}_{ij}=e^{-{\rm i} p\cdot x}\tilde{E}^{\fft{1}{2}(d^2+d-2)}_{ij}\,,
\nn\\
&& f^{{\rm R}(\fft{1}{2}d(d+1))}_{ij}=e^{-{\rm i} p\cdot x}\big(\fft{1}{\ell^2}\tilde{E}^{\fft{1}{2}d(d+1)}_{ij}
-\fft{1}{2r^2}\tilde{E}^{\fft{1}{2}(d^2+d-2)}_{ij}\big)\,.\label{solufrep2}
\eea
In conclusion, the perturbative solutions can be now written in a more compact form
\be
f_{ij}=\sum_I B_I(p,r) \tilde{E}^I_{ij}\,,\label{comp}
\ee
where $B_I(p,r)$ can be read off in (\ref{solufrep2}).  (Note that spin-2 or higher-spin modes in global AdS coordinates can be found in \cite{Chen:2011in,Lu:2011qx}.)

\subsection{Covariant two-point functions}
\label{two-point-result}

\subsubsection{Momentum space}

It follows from (\ref{f-FG}) and (\ref{comp}) that we have, up to some normalization factor (see appendix \ref{B2} for details,) that
\be
f^{(0)}_{ij}=\sum_{I=1}^{\fft{d(d+1)}{2}}A_I(p) \tilde{E}^I_{ij}\,,\qquad f^{(d)}_{ij}= N(d,p)
\sum_{I=d}^{\fft{d^2+d-4}{2}}A_I(p) \tilde{E}^I_{ij}\,,\label{fd-0}
\ee
where the coefficient $N(d,p)$ is given in appendix \ref{B2}; its expression depends on whether $d$ is even or odd. Note we have the orthonormal relation (\ref{orth}), which immediately implies that
\be
A_I(p)=\fft{1}{2}f^{(0)}_{ij}\tilde{E}_I^{ij}\,,\qquad d\leq I\leq \fft{d^2+d-4}{2}\,.
\ee
Therefore, the properly normalized graviton modes in the tensor basis are given by
\bea
 f^{(d)}_{ij}=\ft{1}{2}N(d,p)
\sum_{I=d}^{\fft{d^2+d-4}{2}}\tilde{E}^I_{kl} \tilde{E}^I_{ij}f^{(0)kl}\,.\label{fd}
\eea
It is instructive to introduce the boundary spacetime tensors
\bea
&& \Theta_{ij}(p)=p^2\sum_{k}e^{(k)}_i e^{(k)}_j=\eta_{ij}p^2-p_i p_j\,,
\cr && \Delta^d_{ijkl}(p)=\ft{1}{2}(\Theta_{ik}\Theta_{jl}+\Theta_{il}\Theta_{jk})
-\ft{1}{d-1}\Theta_{ij}\Theta_{kl}\,.
\eea
We can readily verify that
\be
\sum_{I=d}^{\fft{d^2+d-4}{2}}\tilde{E}^I_{kl} \tilde{E}^I_{ij}=\fft{2}{p^4}\Delta^d_{ijkl}(p)\,.
\ee
Consequently, it follows from (\ref{inv-cal}), (\ref{identify}), (\ref{fd}) and the detail expression for $N(d,p)$ given in appendix \ref{B2}, that we have, in momentum space,
\be
\fft{\delta g^{(d)}_{ij}}{\delta g^{(0)kl}}=
\left\{
  \begin{array}{lll}
   -\fft{2^{-d} \pi \ell^{d}\big(-{\rm i}+\cot(\fft{d\pi}{2})\big)}{\Gamma(\fft{d}{2}+1)\Gamma(\fft{d}{2})}
\Delta^d_{ijkl}(p)\,p^{d-4}, & \hbox{\qquad $d$ is odd;} \\
& \\
    -\fft{2^{-d+1}\ell^d}{\Gamma(\fft{d}{2}+1)\Gamma(\fft{d}{2})}\,\Delta^d_{ijkl}(p)\,p^{d-4}\log p, & \hbox{\qquad $d$ is even.}
  \end{array}
\right.
\label{2-pt-momen}
\ee
The structure matches the two-point functions of the energy-momentum tensor of a $d$-dimensional CFT in the momentum space \cite{Coriano:2012wp}.

\subsubsection{Configuration space}

We now transform the two-point functions in momentum space (\ref{2-pt-momen}) into those in configuration space. In the configuration space, we introduce two tensor operators
\bea
\hat{\Theta}_{ij}&=&\partial_i \partial_j-\eta_{ij}\Box\,,\cr
\hat \Delta^d_{ijkl}&=&\ft{1}{2}(\hat\Theta_{ik}\hat\Theta_{jl}+\hat\Theta_{il}\hat\Theta_{jk})
-\ft{1}{d-1}\hat\Theta_{ij}\hat\Theta_{kl}\,.
\eea
We then have
\be
\fft{\delta g^{(d)}_{ij}}{\delta g^{(0)kl}}=
\left\{
  \begin{array}{ll}
-\fft{2^{-d} \pi \ell^{d}(-i+\cot(\fft{d\pi}{2}))}{\Gamma(\fft{d}{2}+1)\Gamma(\fft{d}{2})}\hat{\Delta}^d_{ijkl}
\Big(\int \fft{d^dp}{(2\pi)^d}e^{-{\rm i} p\cdot x}p^{d-4}\Big), & \hbox{\qquad $d$ is odd;} \\
    & \\
  -\fft{2^{-d+1}\ell^d}{\Gamma(\fft{d}{2}+1)\Gamma(\fft{d}{2})}\hat{\Delta}^d_{ijkl}\Big(\int \fft{d^dp}{(2\pi)^d}
e^{-{\rm i} p\cdot x}p^{d-4}\log p\Big), & \hbox{\qquad $d$ is even.}
  \end{array}
\right.
\ee
In order to compute the integrals, we use the formulae
\bea
&&\int \fft{d^dp}{(2\pi)^d}e^{-{\rm i} p\cdot x}p^n=\fft{2^n}{\pi^{\fft{d}{2}}}\fft{\Gamma \left(\frac{d+n}{2}\right)}{\Gamma \left(-\frac{n}{2}\right)}\fft{1}{x^{n+d}}\,,\nn\\
&&\int \fft{d^dp}{(2\pi)^d}e^{-{\rm i} p\cdot x}p^n\log p=
\fft{d}{dn}\int \fft{d^dp}{(2\pi)^d}e^{-{\rm i}p\cdot x}p^n\,.
\eea
Hence, we have
\bea
&& \fft{\delta g^{(d)}_{ij}}{\delta g^{(0)kl}}=\fft{{\rm i} \ell^{d}\Gamma(d-2)}{16\pi^{\fft{d}{2}-1}
\Gamma(\fft{d}{2}+1)\Gamma(\fft{d}{2})\Gamma(2-\fft{d}{2})}\hat{\Delta}^d_{ijkl}
\big(\fft{1}{x^{2(d-2)}}\big)\,,\qquad\qquad \text{$d$ is odd}\,;
\cr && \fft{\delta g^{(d)}_{ij}}{\delta g^{(0)kl}}=-\fft{\ell^{d}\Gamma(0,2-\fft{d}{2})\Gamma(d-2)}{16\pi^{\fft{d}{2}}
\Gamma(\fft{d}{2}+1)\Gamma(\fft{d}{2})\Gamma(2-\fft{d}{2})}\hat{\Delta}^d_{ijkl}
\big(\fft{1}{x^{2(d-2)}}\big)\,,\qquad\qquad \text{$d$ is even}\,.\qquad
\label{2-pt-2}
\eea
Here $\Gamma(0,z)\equiv\Gamma'(z)/\Gamma(z)$ is a digamma function. Evaluating the quantity $\hat{\Delta}^d_{ijkl}(x^{-2(d-2)})$, it turns out that (\ref{2-pt-2}) is equivalent to
\bea
&& \fft{\delta g^{(d)}_{ij}}{\delta g^{(0)kl}}=\fft{{\rm i} (d+1)d(d-2)^2 \Gamma(d-2)}{4\pi^{\fft{d}{2}-1}
\Gamma(\fft{d}{2}+1)\Gamma(\fft{d}{2})\Gamma(2-\fft{d}{2})}\,\fft{\ell^{d}\,\mathcal{I}_{ijkl}(x)}{x^{2d}}
\,,\qquad\qquad\qquad \text{$d$ is odd\,;}\label{2-pt-3}\\
&& \fft{\delta g^{(d)}_{ij}}{\delta g^{(0)kl}}=-\fft{(d+1)d(d-2)^2\Gamma(0,2-\fft{d}{2})\Gamma(d-2)}{4\pi^{\fft{d}{2}}
\Gamma(\fft{d}{2}+1)\Gamma(\fft{d}{2})\Gamma(2-\fft{d}{2})}\,\fft{\ell^{d}\,\mathcal{I}_{ijkl}(x)}{x^{2d}}
\,,\qquad\text{$d$ is even}\,,\nn
\eea
where $\mathcal{I}_{ijkl}(x)$ is given by (\ref{Iijkl}). It is convenient to introduce the purely numerical factors
\bea
N_1 &=&
\left\{
  \begin{array}{ll}
    \fft{\rm i}{2^{d+2} \Gamma(\fft{d}2)^2}, &\qquad\qquad \hbox{$d$ is odd;} \\
&\\
    -\fft{1}{2^{d+1} \pi \Gamma(\fft{d}2)^2}, &\qquad\qquad \hbox{$d$ is even;}
  \end{array}
\right.\nn\\
&&\nn\\
N_2 &=& \fft{\Gamma(d+2)}{8 (-1)^{\fft{d}{2}}(\pi)^{\fft{d}2 +1} (d-1)\Gamma(\fft{d}2)}\,,\qquad\qquad
\hbox{for both even and odd $d$.}\label{n1n2}
\eea
Note that the $(-1)^{d/2}$ in the denominator arises in Lorenzian signature and it is absent in Euclidean signature. The two-point functions can be now compactly written as
\bea
&& \langle T_{ij}(p)T_{kl}(0)\rangle=
\left\{
  \begin{array}{lll}
    N_1\, \mathcal{C}_T\, \Delta^d_{ijkl}(p)p^{d-4}, & \hbox{\qquad$d$ is odd;} \\
    &\\
    N_1\, \mathcal{C}_T\, \Delta^d_{ijkl}(p)p^{d-4}\log p, & \hbox{\qquad $d$ is even,}
  \end{array}
\right.\label{2-pt-universal0}\\
&&\cr
&& \langle T_{ij}(x)T_{kl}(0)\rangle=N_2\, \mathcal{C}_T\, \mathcal{I}_{ijkl}(x)\,{x^{-2d}}\,,\qquad\quad \hbox{for both even and odd $d$,}\label{2-pt-universal}
\eea
where the constant $\mathcal{C}_T$ depends the details of a specific theory. The simplest example is Einstein gravity, and it follows from (\ref{1-pt-Ein}) that we have
\be
\mathcal{C}_T=\ell^{d-1}\,.\label{calCTein}
\ee
Note that in literature, the overall factor $N_2 {\cal C}_T$ of the two-point function in the configuration space
is denoted as $C_T$.  In this paper, we strip off the universal purely numerical factor $N_2$ and introduce the dimensionful quantity ${\cal C}_T$, which is chosen as our convention to have a simple dependence on the AdS radius $\ell$ in the case of Einstein AdS gravity, as in (\ref{calCTein}).

For general massless higher-order gravities with linearized equation (\ref{linear}), we expect that $\mathcal{C}_T=16\pi \kappa_{\rm eff}\ell^{d-1}$. At the first sight, this theory-dependent expression should not be called universal; furthermore, it is expressed in terms of quantities of dual gravity rather than those of the CFT itself.  However, as we discussed in the introduction, we further find a universal expression (\ref{CT-uni}).  Indeed for Einstein gravity, the holographic $a$-charge is
\be
a=\ell^{d-1}\,.
\ee
It follows that the identity (\ref{CT-uni}) is valid straightforwardly.  It should be pointed out that we have also stripped off some inessential overall purely numerical factors of the $a$-charge presented in literature, so that the identity (\ref{CT-uni}) is simple, without encumbered with tedious and inessential numerical numbers.  We use Einstein AdS gravity in general dimensions to set the convention for the identity. It is of interest to test this identity for more general theories.  For general Ricci-polynomial gravities, it was shown \cite{Li:2017ncu} that the decoupling of all the massive modes implies that the linearized theory around the AdS vacua is identical to that of Einstein theory and hence the identity (\ref{CT-uni}) is trivially valid for all these quasi-topological theories. In all these cases, we have a stronger claim that $\mathcal{C}_T=a$.  In the next three sections we shall establish the identity (\ref{CT-uni}) with further nontrivial examples.

\section{Einstein-Gauss-Bonnet Gravity}
\label{inclu-GB}

In this section, we consider Einstein gravity extended with the Gauss-Bonnet term in $D=d+1\ge 5$ dimensions.  For appropriate bare cosmological constant $\Lambda_0=d(d-1)/(2\ell_0^2)$ and the bare coupling constant $\alpha$, the theory admits AdS vacuum of radius $\ell$. For the flat AdS boundary, the total action is given by (see, e.g~\cite{Liu:2008zf},)
\bea
S_{\rm tot} &=& S_{\rm bulk} + S_{\rm GH} + S_{\rm ct}\,,\nn\\
S_{\rm bulk}&=&\fft{1}{16\pi}\int_M d^{d+1}x\sqrt{-g}\Big(R+\fft{d(d-1)}{\ell_0^2}+\alpha(R^2-4R_{\mu\nu}R^{\mu\nu} + R_{\mu\nu\rho\sigma} R^{\mu\nu\rho\sigma})\Big)\,,\nn\\
S_{\rm GH}&=&\fft{1}{8\pi}\int_{\partial M} d^dx\sqrt{-h}\Big(K-\fft{2\alpha}{3}(K^3-3KK^{(2)}+2K^{(3)})\Big)\,,\nn\\
S_{\rm ct}&=&-\fft{1}{8\pi}\int_{\partial M}d^dx \sqrt{-h}\, (d-1)\big(\ell^2-\ft23(d-3)(d-2)\alpha\big)\ell^{-3}\,,\label{egbtotlag}
\eea
where $K^{(2)}\equiv K^i_j K^j_i$, $K^{(3)}\equiv K^i_j K^j_k K^k_i$, and
$(\ell_0,\alpha)$ and $\ell$ are related {\it via} the equations of motion by
\be
\fft{1}{\ell_0^2}-\fft{1}{\ell^2}+\fft{(d-2)(d-3)\alpha}{\ell^4}=0\,.\label{GB-l0}
\ee
The linearized theory around the AdS vacua involves only the graviton (\ref{linear}) with \cite{Fan:2016zfs}
\be
\kappa_{\rm eff}=\fft{1}{16\pi}\big(1-\fft{2\alpha(d-2)(d-3)}{\ell^2}\big)\,.
\ee
Note that in higher-derivative gravities, after setting the Newton's constant to unity, the non-trivial parameters are the bare cosmological constant and bare coupling constants of higher-order curvature invariants such as $\alpha$ for the Gauss-Bonnet term.  When such a theory admits the AdS spacetime of radius $\ell$, these constants and $\ell$ are related by an algebraic equation such as (\ref{GB-l0}) from the equations of motion.
In this paper, we always treat the AdS radius $\ell$ and the bare coupling constants as independent parameters, while solving $\ell_0$ associated with the bare cosmological constant $\Lambda_0$ in terms of these independent parameters $(\ell,\alpha)$ by the equation of motion such as (\ref{GB-l0}).
For the metric ansatz (\ref{FG-co}), the Brown-York energy-momentum tensor is given by \cite{Davis:2002gn}
\bea
&& T_{ij}=-\fft{1}{8\pi}\left(K_{ij}-Kh_{ij}+2\alpha(3J_{ij}-Jh_{ij})
+\fft{(d-1)(3\ell^2-2(d-3)(d-2)\alpha)}{3\ell^3}h_{ij}\right)\,,
\cr && J_{ij}=\fft{1}{3}(2KK_{ik}K_j^k+K_{kl}K^{kl}K_{ij}-2K_{ik}K^{kl}K_{lj}-K^2 K_{ij})\,.\label{BY-GB}
\eea
For the metric (\ref{FG-co}), the extrinsic curvature $K_{ij}$ is given by
\be
K_{ij}=\fft{r}{2\ell}\,\partial_r h_{ij}\,.
\ee
Expanding $K_{ij}$ and $K$ to the relevant order, we have
\be
K_{ij}=\fft{r^2}{\ell^2}g^{(0)}_{ij}-\fft{d-2}{2\ell^2}\fft{1}{r^{d-2}}g^{(d)}_{ij}+\cdots\,,\qquad K=\fft{d}{\ell}+\cdots\,.\label{FG-K}
\ee
Before going on further, it is important to note that ${\rm Tr}g^{(d)}$ is expressed in terms of curvature tensors associated with the boundary \cite{Nojiri:1999mh}, hence we have used the fact that
\be
{\rm Tr}g^{(d)}=0
\ee
for a flat boundary. Analogously, we find
\bea
K_{ik}K_j^{k} &=&\fft{r^2}{\ell^2}g^{(0)}_{ij}-\fft{d-1}{\ell^2}\fft{1}{r^{d-2}}g^{(d)}_{ij}+\cdots\,,\nn\\
K_{kl}K^{kl}K_{ij}
&=&\fft{d}{\ell^3}r^2g^{(0)}_{ij}-\fft{d(d-2)}{\ell^3}\fft{1}{r^{d-2}}g^{(d)}_{ij}+\cdots\,,\nn\\ K_{ik}K^{kl}K_{lj} &=& \fft{r^2}{\ell^3}g^{(0)}_{ij}-\fft{3d-2}{2\ell^3}\fft{1}{r^{d-2}}g^{(d)}_{ij}+\cdots\,.
\eea
With these, we conclude that
\bea
J_{ij} &=& -\fft{(d-2)(d-1)r^2}{3 \ell^3}g^{(0)}_{ij}+\fft{(d-2)(d^2-5 d+2)}{6 \ell^3}\fft{1}{r^{d-2}}g^{(d)}_{ij}+\cdots\,,\nn\\
J &=&-\fft{(d-2)(d-1)d}{3\ell^3}+\cdots\,.\label{FG-J}
\eea
Substituting (\ref{FG-K}), (\ref{FG-J}) and (\ref{FGflat}) back into (\ref{BY-GB}), we arrive at the one-point function
\be
\langle T_{ij}\rangle=\fft{ d}{\ell}\kappa_{\rm eff}\, g^{(d)}_{ij}\,.\label{1-pt-GB}
\ee
Hence, the two-point functions of Einstein-Gauss-Bonnet gravity are given by (\ref{2-pt-universal0}) and (\ref{2-pt-universal}), provided that
\be
\mathcal{C}_T=16\pi \kappa_{\rm eff} \ell^{d-1}=\ell^{d-1}\big(1-\fft{2\alpha(d-3)(d-2)}{\ell^2}\big)\,.
\ee
On the other hand, the $a$-charge of Einstein-Gauss-Bonnet gravity in arbitrary dimensions was previously obtained, given by \cite{Myers:2010xs,Myers:2010tj}
\be
a=\ell^{d-1}\big(1-\fft{2\alpha(d-1)(d-2)}{\ell^2}\big)\,.
\ee
It is remarkably clear that the identity (\ref{CT-uni}) is valid. In particular, when $D=5$, $d=4$, it was shown that $\mathcal{C}_T=c$ \cite{Buchel:2009sk}.  This implies that the $a$-charge and $c$-charge in a four-dimensional CFT can be simply related by
\be
c=\ft13\ell\, \fft{\partial a}{\partial \ell}\,.\label{carelation}
\ee
We shall confirm this identity in massless Einstein-cubic gravities as well in section \ref{cubicgr}.

It is important to note that in this paper we express the holographic $a$ and $c$ charges in terms of the AdS radius $\ell$ and bare coupling constants, such as $\alpha$ for the Gauss-Bonnet term. The relation (\ref{carelation}) and (\ref{CT-uni}) hold only when these parameters are treated as independent.  Our expressions are different from those in literature.  For example, in \cite{Buchel:2009sk,Myers:2010xs,Myers:2010tj}, the charges were expressed in terms of the radius $L$ ($\ell_0$ in our notation) of bare cosmological constant, and it follows that neither the relation (\ref{carelation}) nor (\ref{CT-uni}) were manifest.

\section{Einstein-Lovelock Gravities}
\label{inclu-ll}
The bulk action of general Lovelock gravities \cite{ll} is given by
\be
S_{\text{bulk}}=\fft{1}{16\pi}\int_{M}d^{d+1}x\sqrt{-g}L,\quad \quad L=\sum\limits_k a_k E_k\,,\label{Lovelock}
\ee
where
\be
E_k=\frac{(2k)!}{2^k}\delta^{\mu_1 \mu_2\cdots \mu_{2k}}_{\nu_1 \nu_2 \cdots \nu_{2k}}  R^{\nu_1 \nu_2}_{\mu_1 \mu_2}R^{\nu_3 \nu_4}_{\mu_3 \mu_4} \cdots R^{\nu_{2k-1} \nu_{2k}}_{\mu_{2k-1}\mu_{2k}}\,.
\ee
The multi-index Kronecker delta symbol is defined as follows
\be
\delta^{\mu_1 \mu_2\cdots \mu_{2k}}_{\nu_1 \nu_2 \cdots \nu_{2k}}=\delta^{[\mu_1}_{\nu_1}\delta^{\mu_2}_{\nu_2} \cdots \delta^{\mu_{2k}]}_{\nu_{2k}}\,.
\ee
We further set the bare cosmological constant and Newton's constant by
\be
a_0=\fft{d(d-1)}{\ell_0^2}\,,\qquad a_1=1\,.
\ee
The radius $\ell$ of the AdS vacuum satisfies the algebraic equation \cite{Liu:2017kml}
\be
\sum_{k\geq0}\fft{d!}{(d-2k)!}(-\fft{1}{\ell^2})^k a_k=0\,.
\ee
We find that the effective Newton constant of Lovelock gravity is given by
\be
\kappa_{\rm eff}=\fft{1}{16\pi}\sum_{k\geq1}\fft{(-1)^{k+1}k(d-2)!a_k}{\ell^{2(k-1)}(d-2k)!}\,.
\ee

The Gibbons-Hawking surface term and the counterterms with respect to the flat boundary can be found in \cite{Liu:2017kml}; they are given by
\bea
&& S_{\rm GH}=\fft{1}{16\pi}\int_{\partial M} d^d\sqrt{-h}\sum_{k\geq1}L^{(k)}_{\rm GH}\,,\qquad L^{(k)}_{\rm GH}=\fft{(-1)^{k-1}(2k)!a_k}{2k-1}\delta^{i_1\cdots i_{2k-1}}_{j_1\cdots j_{2k-1}}K_{i_1}^{j_1}\cdots
K_{i_{2k-1}}^{j_{2k-1}}\,,
\cr && S_{\rm ct}=\fft{1}{16\pi}\int_{\partial M} d^dx\sqrt{-h}\sum_{k\geq1}\fft{2k(-1)^k(d-1)!a_k}{(2k-1)(d-2k)!}
(\fft{1}{\ell})^{2k-1}\,.\label{Lov-GB-ct}
\eea
It was shown \cite{Liu:2017kml} that the linearized perturbation of one-point function is
\be
\delta T_{ij}=\sum_{k\geq1}\fft{(-1)^{k}k(d-2)!a_k}{8\pi(d-2k)!}\ell^{2(1-k)}\delta K_{ij}(g)\,,
\ee
where $K_{ij}(g)$ stands for the extrinsic curvature with respect to the metric $g_{ij}$ in (\ref{FG-co}), which can be expanded as
\be
K_{ij}(g)=\fft{r}{2\ell}\partial_r g_{ij}=-\fft{d}{2\ell}g^{(d)}_{ij}+\cdots\,.
\ee
Hence we have
\be
\delta T_{ij}=\sum_{k\geq1}\fft{(-1)^{k+1}k(d-2)!da_k}{16\pi\ell^{2k-1}(d-2k)!}\delta g^{(d)}_{ij}\,.
\ee
Thus we see that the two-point functions for generic Einstein-Lovelock gravities are given by (\ref{2-pt-universal0}) and (\ref{2-pt-universal}) with
\be
\mathcal{C}_T=16\pi \kappa_{\rm eff} \ell^{d-1}=\ell^{d-1}\sum_{k\geq1}\fft{(-1)^{k+1}k(d-2)!}{\ell^{2(k-1)}(d-2k)!}a_k\,.
\ee

To obtain the $a$-charge of Lovelock gravity, we employ the trick of the reduced FG expansions developed in \cite{Li:2017txk,Li:2018kqp}.  (See also, e.g.~\cite{Alkac:2018whk}). Considering a special class of the FG coordinates
\be
ds_D^2=\fft{\ell^2}{4r^2}dr^2+\fft{f(r)}{r}d\Omega_{d}^2\,,
\ee
where the AdS boundary is located at $r=0$, we find
\bea
&& R_{ir}^{jr}=F_1\delta_i^j\,,\qquad R_{ij}^{kl}=F_2(\delta_i^k \delta_j^l-\delta_j^k \delta_i^l)\,,
 \qquad F_1=-\fft{2 r^2 f f''-r^2f'^2+f^2}{\ell^2 f^2}\,,
  \cr && F_2=-\fft{-r f (\ell^2+2 f')+r^2 f'^2+f^2}{\ell^2 f^2}\,.
\eea
The bulk action is then given by
\be
S_{\rm bulk}=\fft{1}{16\pi}\int d^{d+1}x\sqrt{-g}\sum_k a_k\Big(\fft{d!}{(d-2k)!}F_2^k+\fft{2k\, d!}{(d-2k+1)!}F_2^{k-1}F_1\Big)\,.
\ee
We expand $f$ as
\be
f=f_0+f_2 r+f_4 r^2+f_6 r^3+\cdots\,.
\ee
Following the procedure developed in\cite{Li:2017txk,Li:2018kqp}, we have
\be
f_2=-\fft{\ell^2}{2}\,,\qquad f_4=\fft{\ell^4}{16f_0}\,,\qquad f_{2i}=0\,,\qquad i\geq 3\,.
\ee
We can then read off the $a$-charge:
\be
a=\ell^{d-1}\sum_{k\geq1}\fft{(-1)^{k+1}k(d-2)!}{\ell^{2k}(d-2k+1)!} a_k\,.
\ee
It is easy to verify that (\ref{CT-uni}) is again valid.

\section{Einstein-Riemann Cubic Gravities}
\label{cubicgr}

The crucial property in our derivation is that the linearized spectrum of the AdS vacuum involves only the massless graviton, governed by the linearized equations of motion (\ref{linear}).  Einstein gravity, Einstein-Guass-Bonnet and Einstein-Lovelock gravities clearly all have this property.  In fact these theories are of two derivatives in any background.  For AdS vacua, the condition can be relaxed and in this section, we consider Riemann cubic extended gravities.

\subsection{The  action}
The bulk action of Einstein gravity extended with generic Riemann cubic invariants is
\be
S=\fft{1}{16\pi}\int_{M} d^{d+1}x\sqrt{-g}L\,,\qquad L=R+\fft{d(d-1)}{\ell_0^2}+H^{(3)}\,,\label{bulk}
\ee
where $H^{(3)}$ is given by
\bea
 H^{(3)}&=& e_1 R^3 + e_2 R\,R_{\mu\nu} R^{\mu\nu} + e_3 R^{\mu}_{\nu} R^{\nu}_\rho R^{\rho}_\mu + e_4 R^{\mu\nu} R^{\rho\sigma} R_{\mu\rho\nu\sigma}\cr
&&+ e_5 R R^{\mu\nu\rho\sigma} R_{\mu\nu\rho\sigma} +e_6 R^{\mu\nu} R_{\mu \alpha\beta\gamma} R_{\nu}{}^{\alpha\beta\gamma} +
e_7 R^{\mu\nu}{}_{\rho\sigma} R^{\rho\sigma}{}_{\alpha\beta} R^{\alpha\beta}{}_{\mu\nu}\cr
&&+e_8 R^\mu{}_\nu{}^\alpha{}_\beta R^\nu{}_\rho{}^\beta{}_\gamma R^{\rho}{}_\mu{}^\gamma{}_{\alpha}\,.\label{riemanncubiclag}
\eea
The bare and effective cosmological constants are related by \cite{Li:2017txk}
\bea
&& \fft{1}{\ell_0^2}-\fft{1}{\ell^2}- \fft{(d-5)}{(d-1)\ell^6}
\Big(d^2(d+1)^2 e_1+d^2(d+1) e_2+d^2 e_3\cr
&&+d^2 e_4+2 d(d+1) e_5+2 d e_6+4 e_7+(d-1) e_8
\Big)=0\,.
\eea
It is necessary to add the Gibbons-Hawking surface term to make the variation principle well defined. To do this, we define
\be
P_{\mu\nu\rho\sigma}\equiv\fft{\partial L}{\partial R^{\mu\nu\rho\sigma}}=P^0_{\mu\nu\rho\sigma}+\sum_{i=1}^8 e_i P^{i}_{\mu\nu\rho\sigma}\,,
\ee
where
\bea
P^0_{\mu\nu\rho\sigma} &=& \ft12(g_{\mu\rho} g_{\nu\sigma} - g_{\mu\sigma} g_{\nu\rho})\,,\qquad
P^1_{\mu\nu\rho\sigma} = \ft32(g_{\mu\rho} g_{\nu\sigma} - g_{\mu\sigma} g_{\nu\rho}) R^2\,,\cr
P^2_{\mu\nu\rho\sigma} &=& \ft12  (g_{\mu\rho} g_{\nu\sigma} - g_{\mu\sigma} g_{\nu\rho}) R_{\alpha\beta}R^{\alpha\beta} \cr
&& + \ft12 R\, (g_{\mu\rho} R_{\nu\sigma} - g_{\mu\sigma} R_{\nu\rho}
-g_{\nu\rho} R_{\mu\sigma} + g_{\nu\sigma} R_{\mu\rho})\,,\cr
P^3_{\mu\nu\rho\sigma} &=& \ft34  \big(g_{\mu\rho} R_{\nu\gamma}R_{\sigma}{}^\gamma - g_{\mu\sigma} R_{\nu\gamma} R_{\rho}{}^\gamma  -
g_{\nu\rho} R_{\mu\gamma}R_{\sigma}{}^\gamma + g_{\nu\sigma} R_{\mu\gamma} R_{\rho}{}^\gamma\big)\,,\cr
P^4_{\mu\nu\rho\sigma}&=&\ft12(g_{\nu\sigma}R_{\mu\alpha\rho\beta}-g_{\nu\rho}R_{\mu\alpha\sigma\beta}-
g_{\mu\sigma}R_{\nu\alpha\rho\beta}+g_{\mu\rho}R_{\nu\alpha\sigma\beta})\cr
&&+\ft{1}{2} (R_{\mu\rho}R_{\nu\sigma}-R_{\mu\sigma}R_{\nu\rho}+R^{\alpha\beta})\,,\cr
P^5_{\mu\nu\rho\sigma}&=& \ft12(g_{\mu\rho} g_{\nu\sigma} - g_{\mu\sigma} g_{\nu\rho})R^{\alpha\beta\gamma\eta}R_{\alpha\beta\gamma\eta}+2RR_{\mu\nu\rho\sigma}\,,\cr
P^6_{\mu\nu\rho\sigma}&=&\ft{1}{2}(R_{\nu}^{\alpha}R_{\mu\alpha\rho\sigma}+
R_{\sigma}^{\alpha}R_{\mu\nu\rho\alpha}
-R_{\rho}^{\alpha}R_{\mu\nu\sigma\alpha}-R_{\mu}^{\alpha}R_{\nu\alpha\rho\sigma})\cr
&&+\ft14(g_{\nu\sigma}R_{\mu}^{\alpha\beta\gamma}-g_{\mu\sigma}R_{\nu}^{\alpha\beta\gamma})
R_{\rho\alpha\beta\gamma}
+\ft14(g_{\mu\rho}R_{\nu}^{\alpha\beta\gamma}-g_{\nu\rho}R_{\mu}^{\alpha\beta\gamma})
R_{\sigma\alpha\beta\gamma}
\,,\cr
P^7_{\mu\nu\rho\sigma}&=&3 R_{\mu\nu}^{\alpha\beta}R_{\rho\sigma\alpha\beta}\,,\qquad
P^8_{\mu\nu\rho\sigma}=\ft32  (R_{\mu}{}^{\alpha}{}_{\rho}{}^{\beta} R_{\nu\alpha\sigma\beta}-R_{\mu}{}^{\alpha}{}_{\sigma}{}^{\beta} R_{\nu\alpha\rho\beta})\,.
\eea
The surface term is then given by \cite{Deruelle:2009zk}
\be
S_{\rm GH}=\fft{1}{4\pi}\int_{\partial M} d^dx\sqrt{-h} \Phi^\mu_\nu K_{\mu}^{\nu}\,,\qquad
\Phi^\mu_\nu=P^\mu{}_{\rho\nu\sigma} n^\rho n^\sigma\,,\label{surf}
\ee
where $n^\mu$, in our coordinate choice (\ref{FG-co}), is given by
\be
n^\mu=\fft{r}{\ell}(\fft{\partial}{\partial r})^\mu\,.
\ee
It is important to note that in this description, $\Phi^\mu_\nu$ is auxiliary and does not involve in the variation of $\delta g_{\mu\nu}$.
We shall also need to include the appropriate counterterm so as to make the on-shell action finite. After some algebra, we find that the counterterm associated with the flat boundary is given by
\bea
S_{\rm ct}&=&-\fft{1}{8\pi}\int d^dx\sqrt{-h}\,\fft{3(d-1)}{\ell^5}\Big(\ft13\ell^4+d^2(d+1)^2 e_1+d^2(d+1) e_2\nn\\
&&\qquad +d^2 e_3+d^2 e_4+2 d(d+1)\, e_5+2 d\, e_6+4 e_7+(d-1) e_8\Big)\,.\label{ct}
\eea
As we shall discuss in the next subsection, for appropriate coupling constants, the cubic theory becomes massless gravity where the massive scalar and spin-2 modes are decoupled.  In this case, we can introduce the Gibbons-Hawking surface term without needing the auxiliary field.  We find
\be
S_{\rm GH}=\fft{1}{4\pi}\int_{\partial M} d^dx\sqrt{-h}\, \widetilde P_{\mu\nu\rho\sigma}\, K^{\mu\rho} \, n^\nu n^\sigma\,,\label{surf1}
\ee
where
\be
\widetilde P_{\mu\nu\rho\sigma} = P^0_{\mu\nu\rho\sigma} + \fft15\sum_{i=1}^8 e_i P^i_{\mu\nu\rho\sigma}\,.
\ee
The variation of the action with respect to $\delta g_{\mu\nu}$ now requires varying all the fields in the integrand.  The ``$1/5$'' factor in the surface term is analogous to the ``$1/3$'' factor in the Gibbons-Hawking term in Einstein Gauss-Bonnet gravity (\ref{egbtotlag}). Consequently, we now require
a different counterterm, given by
\bea
S_{\rm ct}&=&-\fft{1}{8\pi}\int d^dx\sqrt{-h}\,\fft{3(d-5)}{5\ell^5}\Big(\fft{5(d-1)}{3(d-5)}\ell^4+d^2(d+1)^2 e_1+d^2(d+1) e_2\nn\\
&&\qquad+d^2 e_3+d^2 e_4+2 (d+1) d\, e_5+2 d\, e_6+4 e_7+(d-1) e_8\Big)\,.\label{ct1}
\eea
It should be emphasized that the boundary terms (\ref{surf}) and (\ref{ct}) are generally valid; on the other hand, the new boundary terms (\ref{surf1}) and (\ref{ct1}), which are more convenient for performing variations, are valid only for massless cubic gravities. As we shall see in the next subsection, for massless cubic gravities, the Brown-York energy-momentum tensors associated with the linear perturbations around the AdS vacuum can be derived from the two seemingly different boundary terms, and the results turn out to be identically the same.

\subsection{Two-point functions}

The decoupling of both the massive scalar and spin-2 modes requires, (see e.g.~\cite{Sisman:2011gz,Bueno:2016xff,Li:2017txk},)
\bea
&&(d+1)d e_2+3d e_3+(2d-1) e_4+4(d+1)d e_5+4 (d+1) e_6+24 e_7-3 e_8=0\,,\nn\\
&&12 (d+1) d^2 e_1+\left(d^2+10d+1\right) d e_2+3 (d+1) d e_3+\left(2 d^2+5d-1\right) e_4\nn\\
&&+4 (d+5) d e_5+4 (2 d+1) e_6+3 (d-1) e_8+24 e_7=0\,.\label{ghost-free}
\eea
The remaining six-parameter theory is what we call massless cubic gravity and the linearized equation of motion around the AdS vacuum is given by (\ref{linear}) with
\be
\kappa_{\rm eff}=\fft{1}{16\pi}\Big(1+\fft{1}{\ell^4}(d-5)(d-2)(3(d+1)de_1+2de_2+e_4+4e_5)\Big)\,.
\ee
Further specialization of the parameters can yield quasi-topological gravity \cite{Oliva:2010eb,Myers:2010ru,Myers:2010jv}, Einsteinian cubic gravity \cite{Bueno:2016xff,Bueno:2018xqc} or quasi-topological Ricci polynomial gravity \cite{Li:2017ncu}.

For this generic massless cubic gravity, a covariant Brown-York energy-momentum tensor is not available in the literature. We instead adopt the second equation in (\ref{1-pt-0}) to derive the one-point function. We substitute the flat FG expansion (\ref{FGflat}) into the total action with (\ref{bulk}), (\ref{surf}) and (\ref{ct}). We then perform the variation associated with $g^{(0)ij}$, keeping in mind that $\Phi^\mu_\nu$ should be left unvaried, we arrive at the one-point function:
\be
\langle T_{ij}\rangle=\fft{d}{\ell} \kappa_{\rm eff}\, g^{(d)}_{ij}\,.
\ee
Alternatively, we use the total action involving (\ref{bulk}), (\ref{surf1}) and (\ref{ct1}) and vary all the fields associated with $g^{(0)ij}$, and we find that the result turns out to be identically the same.

Having obtained the one-point function, it follows from our earlier discussions that the two-point functions are given by (\ref{2-pt-universal}), with the coefficient $\mathcal{C}_T$
\bea
\mathcal{C}_T&=&16\pi \kappa_{\rm eff} \ell^{d-1}\nn\\
&=&\ell^{d-1}\Big(1+\fft{1}{\ell^4}(d-5)(d-2)(3(d+1)de_1+2de_2+e_4+4e_5)\Big)\,.
\eea
On the other hand, the holographic $a$-charge for the massless cubic theory in arbitrary dimensions is \cite{Li:2017txk}
\be
a=\ell^{d-1}\Big(1+\fft{1}{\ell^4}(d-2)(d-1)(3(d+1)de_1+2de_2+e_4+4e_5)\Big)\,,
\ee
Thus we see that the relation (\ref{CT-uni}) is indeed valid.

It is also intriguing to note that in $D=5$ ($d=4$), the $c$-charge for the massless cubic theory is \cite{Li:2017txk}
\be
c=\ell^3 - 2(60 e_1 + 8 e_2 + e_4 + 4 e_5)\ell^{-1}\,.
\ee
Thus in $d=4$, we have $\mathcal{C}_T=c$, and the relation (\ref{carelation}) is again established.
Recalling also the remarks at the end of section \ref{inclu-GB} for the Einstein-Gauss-Bonnet theory, we have sufficient evidence to conjecture that the relation (\ref{carelation}) between $c$ and $a$ charges is a general property of CFTs in four dimensions.

\section{Conclusion}
\label{conc}

In this paper, we considered Einstein gravity extended with general classes of higher-order curvature invariants.  For appropriate coupling constants, these theories admit the AdS vacuum of certain radius $\ell$. We restrict to massless higher-order gravities where the linearized spectrum around the AdS vacuum involves only the massless graviton, satisfying the linearized equation (\ref{linear}), where the effective Newton's constant $1/\kappa_{\rm eff}$ is modified by the higher-order contributions.

We derived the covariant holographic two-point functions of these pure gravity theories in the AdS vacuum. We presented the results in both momentum and configuration spaces.  We first demonstrated that the two-point functions can be organized in a universal form (\ref{2-pt-uni}) up to a theory-dependent constant $\mathcal{C}_T$.  We found that $\mathcal{C}_T=16\pi \kappa_{\rm eff} \ell^{d-1}$, which is not surprising. However, we further found that this quantity was related to the holographic $a$-charge by a universal expression
\be
\mathcal{C}_T= \fft{1}{d-1} \ell \fft{\partial a}{\partial \ell}\,.\label{relation}
\ee
We verified this relation, using large classes of massless high-order gravities, including quasi-topological Ricci polynomial,  Einstein-Gauss-Bonnet, Einstein-Lovelock and Einstein-Riemann cubic gravities. It should be understood that the $a$-charge in (\ref{relation}) is given as function of the AdS radius $\ell$ and bare coupling constants of higher-order curvature invariants as independent variables, whilst the bare cosmological constant is solved in terms of these variables by the equations of motion.

Another intriguing observation is that the holographic $c$ and $a$ charges in $d=4$ are related by (\ref{carelation}).  It is of great interest to investigate the higher-dimensional generalization of this relation.
Since we derived the relations using the holographic technique, it is of great importance to test these holographic conclusions directly in conformal field theories.

\section*{Acknolwedgement}

We are grateful to Zhao-Long Wang for useful discussions. The work is supported in part by
NSFC grants No.~11475024 and No.~11875200.

\appendix

\section{Asymptotic Behavior of Hankel Functions}
\label{B2}

The two-point functions we obtained in this paper involve the linear graviton modes, the overall coefficient thus depends on the normalization of these modes.  It follows from the discussions in section \ref{cov-stru}, we need to the normalize the function $r^{-\fft{d}{2}}H_{\fft{d}{2}}(\fft{|p|\ell}{r})$. To be specific, we introduce a normalization factor $C$ such that
\be
\lim_{r\rightarrow\infty}C\,r^{-\fft{d}{2}} H_{\fft{d}{2}}(\fft{|p|\ell}{r})=1\,.\label{normalhankel}
\ee
The asymptotic behavior of $r^{-\fft{d}{2}} H_{\fft{d}{2}}(\fft{|p|\ell}{r})$ is
\bea
r^{-\fft{d}{2}} H_{\fft{d}{2}}(\fft{|p|\ell}{r})&=&-\fft{{\rm i} 2^{d/2} \Gamma (\fft{d}{2}) (\ell p)^{-\fft{d}{2}}}{\pi }-\fft{{\rm i} 2^{\fft{d}{2}-1} \Gamma (\frac{d}{2}) (\ell p)^{2-\fft{d}{2}}}{\pi  (d-2) r^2}-\fft{{\rm i} 2^{\fft{d}{2}-3} \Gamma (\frac{d}{2}) (\ell p)^{-\fft{d}{2}}}{\pi  (d-4)(d-2) r^4}
\cr && +\cdots+\fft{2^{-\fft{d}{2}} (\ell p)^{d/2} (\pi -{\rm i} \cos(\fft{\pi  d}{2}) \Gamma (\frac{d}{2}+1) \Gamma(-\frac{d}{2}))}{\pi  \Gamma(\frac{d}{2}+1)}\fft{1}{r^d}+\cdots\,.
\eea
The normalization factor $C$ can be easily determined, given by
\be
C=\fft{{\rm i} \pi (\ell p)^{\fft{d}{2}}}{ 2^{d/2} \Gamma (\fft{d}{2})}\,.
\ee
However, there are some subtleties that are worth discussing here. The normalized function approaches the AdS boundary as
\bea
C\,r^{-\fft{d}{2}} H_{\fft{d}{2}}(\fft{|p|\ell}{r})&=&1+\fft{\ell^2 p^2}{2(d-2)r^2}+\fft{\ell^4 p^4}{8(d-4)(d-2)r^4}+\cdots\nn\\
 &&-\fft{2^{-d} \pi\ell^d(-i+\cot(\fft{d\pi}{2}))}{\Gamma(\fft{d}{2}+1)\Gamma(\fft{d}{2})}\fft{p^d}{r^d}+\cdots\,.
\label{hasym}
\eea
Thus when $d$ is odd, we can simply read off the coefficient in (\ref{fd-0}):
\be
N(d,p)=\fft{{\rm i} 2^{-d} \pi\ell^d}{\Gamma(\fft{d}{2}+1)\Gamma(\fft{d}{2})}p^d\,,\qquad \text{$d=$ odd}\,.
\ee
Note that $\cot(d Pi/2)=0$ for odd $d$. The situation becomes more complicated when $d=2k$ is even.  Firstly the above expression is divergent. Secondly, an additional term of order $1/r^d$ emerges in the ellipses in the first line of (\ref{hasym}) and it diverges too! After using the dimensional regularization scheme $d=2k+\epsilon$, this new term is given by
\be
\fft{1}{2^{k} k!}\fft{(d-2k-1)!}{(d-2)!}\fft{(\ell p)^{2k}}{r^{2k}}=
\fft{1}{2^{2k-1}k!(k-1)!\epsilon}\fft{(\ell p)^{2k}}{r^{2k}}-\fft{1}{2^{2k}k!(k-1)!}
(\gamma+\Gamma(0,k))\fft{(\ell p)^{2k}}{r^{2k}}\,.\label{add}
\ee
We can now easily verify that the divergent part of (\ref{add}) cancels exactly the divergent part in the second line of (\ref{hasym}), i.e
\be
-\fft{2^{-d} \pi\ell^d(-{\rm i}+\cot(\fft{d\pi}{2}))}{\Gamma(\fft{d}{2}+1)\Gamma(\fft{d}{2})}\fft{p^d}{r^d}
=-\fft{1}{2^{2k-1}k!(k-1)!\epsilon}\fft{(\ell p)^{2k}}{r^{2k}}-\fft{1}{2^{2k-1}k!(k-1)!}\fft{(\ell p)^{2k}\log p}{r^{2k}}+\cdots\,,\label{log}
\ee
where the dots represent further finite terms with the factor of $\fft{p^{2k}}{r^{2k}}$.   Note that in even $d=2k$ dimensions, terms of $p^{n}$ for $n\in\mathbb{Z^+}$ contribute nothing to the two-point functions, since they are the contact terms contributing $\Box^n \delta(x)$. The relevant nontrivial term of order $1/r^{2k}$ thus involves $p^{2k}\log p$ only, from the second line of (\ref{hasym}).  Thus we have
\be
N(d,p)=-\fft{2^{-d+1}}{\Gamma(\fft{d}{2}+1)\Gamma(\fft{d}{2})}(\ell p)^{d}\log p\,,\qquad \text{$d=$ even}\,.
\ee


\begin{thebibliography}{99}

\bibitem{Maldacena:1997re}
  J.M.~Maldacena,
  ``The Large $N$ limit of superconformal field theories and supergravity,''
  Int.\ J.\ Theor.\ Phys.\  {\bf 38}, 1113 (1999)
  [Adv.\ Theor.\ Math.\ Phys.\  {\bf 2}, 231 (1998)]
  doi:10.1023/A:1026654312961
  [hep-th/9711200].

\bibitem{deBoer:2000cz}
J.~de Boer,
``The holographic renormalization group,''
Fortsch.\ Phys.\  {\bf 49} (2001) 339
doi: 10.1002/1521-3978(200105)49:4/6(339::AID-PROP339)3.0.CO;2-A
[hep-th/0101026];


\bibitem{Bianchi:2001kw}
M.~Bianchi, D.Z.~Freedman and K.~Skenderis,
``Holographic renormalization,''
Nucl.\ Phys.\ B {\bf 631} (2002) 159
doi:10.1016/S0550-3213(02)00179-7
[hep-th/0112119].


\bibitem{Skenderis:2002wp}
K.~Skenderis,
``Lecture notes on holographic renormalization,''
Class.\ Quant.\ Grav.\  {\bf 19} (2002) 5849
doi:10.1088/0264-9381/19/22/306
[hep-th/0209067].


\bibitem{Gubser:1998bc}
S.S.~Gubser, I.R.~Klebanov and A.M.~Polyakov,
``Gauge theory correlators from noncritical string theory,''
Phys.\ Lett.\ B {\bf 428} (1998) 105
doi:10.1016/S0370-2693(98)00377-3
[hep-th/9802109].

\bibitem{Witten:1998qj}
E.~Witten,
``Anti-de Sitter space and holography,''
Adv.\ Theor.\ Math.\ Phys.\  {\bf 2} (1998) 253
doi:10.4310/ATMP.1998.v2.n2.a2
[hep-th/9802150].

\bibitem{Freedman:1998tz}
D.Z.~Freedman, S.D.~Mathur, A.~Matusis and L.~Rastelli,
``Correlation functions in the CFT$_d$/AdS$_{d+1}$ correspondence,''
Nucl.\ Phys.\ B {\bf 546} (1999) 96
doi:10.1016/S0550-3213(99)00053-X
[hep-th/9804058].

\bibitem{Muck:1998rr}
W.~Mueck and K.S.~Viswanathan,
``Conformal field theory correlators from classical scalar field theory on AdS$_{d+1}$,''
Phys.\ Rev.\ D {\bf 58} (1998) 041901
doi:10.1103/PhysRevD. 58.041901
[hep-th/9804035].

\bibitem{Liu:1998bu}
H.~Liu and A.A.~Tseytlin,
``$D = 4$ superYang-Mills, $D = 5$ gauged supergravity, and $D = 4$ conformal supergravity,''
Nucl.\ Phys.\ B {\bf 533} (1998) 88
doi:10.1016/S0550-3213 (98)00443-X
[hep-th/9804083].

\bibitem{KeskiVakkuri:1998nw}
E.~Keski-Vakkuri,
``Bulk and boundary dynamics in BTZ black holes,''
Phys.\ Rev.\ D {\bf 59} (1999) 104001
doi:10.1103/PhysRevD.59.104001
[hep-th/9808037].

\bibitem{Stelle:1976gc}
  K.S.~Stelle,
``Renormalization of higher derivative quantum gravity,''
  Phys.\ Rev.\ D {\bf 16}, 953 (1977).
  doi:10.1103/PhysRevD.16.953

\bibitem{Lu:2011zk}
  H.~L\"u and C.N.~Pope,
``Critical gravity in four dimensions,''
  Phys.\ Rev.\ Lett.\  {\bf 106}, 181302 (2011)
  doi:10.1103/PhysRevLett.106.181302
  [arXiv:1101.1971 [hep-th]].

\bibitem{Deser:2011xc}
  S.~Deser, H.~Liu, H.~L\"u, C.N.~Pope, T.C.~Sisman and B.~Tekin,
``Critical points of $D$-dimensional extended gravities,''
  Phys.\ Rev.\ D {\bf 83}, 061502 (2011)
  doi:10.1103/Phys RevD.83.061502
  [arXiv:1101.4009 [hep-th]].

\bibitem{Johansson:2012fs}
N.~Johansson, A.~Naseh and T.~Zojer,
``Holographic two-point functions for $4d$ log-gravity,''
JHEP {\bf 1209} (2012) 114
doi:10.1007/JHEP09(2012)114
[arXiv:1205.5804 [hep-th]].

\bibitem{Ghodsi:2014hua}
A.~Ghodsi, B.~Khavari and A.~Naseh,
``Holographic two-point functions in conformal gravity,''
JHEP {\bf 1501} (2015) 137
doi:10.1007/JHEP01(2015)137
[arXiv:1411.3158 [hep-th]].

\bibitem{ll}
  D.~Lovelock,
``The Einstein tensor and its generalizations,''
  J.\ Math.\ Phys.\  {\bf 12}, 498 (1971).

\bibitem{Tekin:2016vli}
B.~Tekin,
``Particle content of quadratic and $f(R_{\mu\nu\sigma \rho})$ theories in $(A)dS$,''
Phys.\ Rev.\ D {\bf 93} (2016) no.10,  101502
doi:10.1103/PhysRevD.93.101502
[arXiv:1604.00891 [hep-th]].

\bibitem{Bueno:2016xff}
P.~Bueno and P.A.~Cano,
``Einsteinian cubic gravity,''
Phys.\ Rev.\ D {\bf 94} (2016) no.10,  104005
doi:10.1103/PhysRevD.94.104005
[arXiv:1607.06463 [hep-th]].


\bibitem{Bueno:2016ypa}
P.~Bueno, P.A.~Cano, V.S.~Min and M.R.~Visser,
``Aspects of general higher-order gravities,''
Phys.\ Rev.\ D {\bf 95} (2017) no.4,  044010
doi:10.1103/PhysRevD.95.044010
[arXiv:1610.08519 [hep-th]].

\bibitem{Li:2017ncu}
Y.Z.~Li, H.S.~Liu and H.~L\"u,
``Quasi-topological Ricci polynomial gravities,''
JHEP {\bf 1802} (2018) 166
doi:10.1007/JHEP02(2018)166
[arXiv:1708.07198 [hep-th]].

\bibitem{Dehghani:2011vu}
  M.H.~Dehghani, A.~Bazrafshan, R.B.~Mann, M.R.~Mehdizadeh, M.~Ghanaatian and M.H.~Vahidinia,
``Black holes in quartic quasitopological gravity,''
Phys.\ Rev.\ D {\bf 85} (2012) 104009
doi:10.1103/PhysRevD.85.104009
[arXiv:1109.4708 [hep-th]].


\bibitem{Karasu:2016ifk}
  A.~Karasu, E.~Kenar and B.~Tekin,
``Minimal extension of Einstein¡¯s theory: The quartic gravity,''
Phys.\ Rev.\ D {\bf 93} (2016) no.8,  084040
doi:10.1103/PhysRevD.93.084040
[arXiv:1602.02567 [hep-th]].

\bibitem{Bueno:2016dol}
  P.~Bueno, P.A.~Cano, A.O.~Lasso and P.F.~Ram¨ªrez,
``f(Lovelock) theories of gravity,''
JHEP {\bf 1604} (2016) 028
doi:10.1007/JHEP04(2016)028
[arXiv:1602.07310 [hep-th]].

\bibitem{Cisterna:2017umf}
  A.~Cisterna, L.~Guajardo, M.~Hassaine and J.~Oliva,
``Quintic quasi-topological gravity,''
JHEP {\bf 1704} (2017) 066
doi:10.1007/JHEP04(2017)066
[arXiv:1702.04676 [hep-th]].

\bibitem{Hennigar:2017ego}
  R.A.~Hennigar, D.~Kubiz\v n\'ak and R.B.~Mann,
``Generalized quasitopological gravity,''
Phys.\ Rev.\ D {\bf 95} (2017) no.10,  104042
doi:10.1103/PhysRevD.95.104042
[arXiv: 1703.01631 [hep-th]].


\bibitem{Ahmed:2017jod}
  J.~Ahmed, R.A.~Hennigar, R.B.~Mann and M.~Mir,
``Quintessential quartic quasi-topological quartet,''
JHEP {\bf 1705} (2017) 134
doi:10.1007/JHEP05(2017)134
[arXiv: 1703.11007 [hep-th]].

\bibitem{Feng:2018qnx}
  X.H.~Feng, H.~Huang, S.L.~Li, H.~L\"u and H.~Wei,
  ``Cosmological time crystals from Einstein-Cubic gravities,''
  arXiv:1807.01720 [hep-th].


\bibitem{Osborn:1993cr}
H.~Osborn and A.~C.~Petkou,
``Implications of conformal invariance in field theories for general dimensions,''
Annals Phys.\  {\bf 231} (1994) 311
doi:10.1006/aphy.1994.1045
[hep-th/9307010].

\bibitem{Erdmenger:1996yc}
J.~Erdmenger and H.~Osborn,
``Conserved currents and the energy momentum tensor in conformally invariant theories for general dimensions,''
Nucl.\ Phys.\ B {\bf 483} (1997) 431
doi:10.1016/S0550-3213(96)00545-7
[hep-th/9605009].

\bibitem{Coriano:2012wp}
C.~Coriano, L.~Delle Rose, E.~Mottola and M.~Serino,
``Graviton vertices and the mapping of anomalous correlators to momentum space for a general conformal field theory,''
JHEP {\bf 1208} (2012) 147
doi:10.1007/JHEP08(2012)147
[arXiv:1203.1339 [hep-th]].

\bibitem{Buchel:2009sk}
A.~Buchel, J.~Escobedo, R.C.~Myers, M.F.~Paulos, A.~Sinha and M.~Smolkin,
``Holographic GB gravity in arbitrary dimensions,''
JHEP {\bf 1003} (2010) 111
doi:10.1007/ JHEP03(2010)111
[arXiv:0911.4257 [hep-th]].

\bibitem{Li:2017txk}
Y.Z.~Li, H.~L\"u and J.B.~Wu,
``Causality and $a$-theorem constraints on Ricci polynomial and Riemann cubic gravities,''
Phys.\ Rev.\ D {\bf 97} (2018) no.2,  024023
doi:10.1103/Phys RevD.97.024023
[arXiv:1711.03650 [hep-th]].


\bibitem{Henningson:1998gx}
M.~Henningson and K.~Skenderis,
``The holographic Weyl anomaly,''
JHEP {\bf 9807} (1998) 023
doi:10.1088/1126-6708/1998/07/023
[hep-th/9806087];


\bibitem{Henningson:1998ey}
M.~Henningson and K.~Skenderis,
``Holography and the weyl anomaly,''
Fortsch.\ Phys.\  {\bf 48} (2000) 125
[hep-th/9812032].

\bibitem{Imbimbo:1999bj}
C.~Imbimbo, A.~Schwimmer, S.~Theisen and S.~Yankielowicz,
``Diffeomorphisms and holographic anomalies,''
Class.\ Quant.\ Grav.\  {\bf 17} (2000) 1129
doi:10.1088/0264-9381/ 17/5/322
[hep-th/9910267].


\bibitem{Nojiri:1999mh}
S.~Nojiri and S.D.~Odintsov,
``On the conformal anomaly from higher derivative gravity in AdS/CFT correspondence,''
Int.\ J.\ Mod.\ Phys.\ A {\bf 15} (2000) 413
doi:10.1142/ S0217751X00000197
[hep-th/9903033].

\bibitem{Blau:1999vz}
M.~Blau, K.S.~Narain and E.~Gava,
``On subleading contributions to the AdS/CFT trace anomaly,''
JHEP {\bf 9909} (1999) 018
doi:10.1088/1126-6708/1999/09/018
[hep-th/9904179].


\bibitem{Myers:2010xs}
R.C.~Myers and A.~Sinha,
``Seeing a $c$-theorem with holography,''
Phys.\ Rev.\ D {\bf 82} (2010) 046006
doi:10.1103/PhysRevD.82.046006
[arXiv:1006.1263 [hep-th]].

\bibitem{Myers:2010tj}
R.C.~Myers and A.~Sinha,
``Holographic $c$-theorems in arbitrary dimensions,''
JHEP {\bf 1101} (2011) 125
doi:10.1007/JHEP01(2011)125
[arXiv:1011.5819 [hep-th]].

\bibitem{deHaro:2000vlm}
S.~de Haro, S.N.~Solodukhin and K.~Skenderis,
``Holographic reconstruction of space-time and renormalization in the AdS/CFT correspondence,''
Commun.\ Math.\ Phys.\  {\bf 217} (2001) 595
doi:10.1007/s002200100381
[hep-th/0002230].

\bibitem{Fischetti:2012rd}
D.~Marolf, W.~Kelly and S.~Fischetti,
``Conserved charges in asymptotically (locally) AdS spacetimes,''
doi:10.1007/978-3-642-41992-8-19
arXiv:1211.6347 [gr-qc].

\bibitem{Chen:2011in}
  Y.X.~Chen, H.~L\"u and K.N.~Shao,
  ``Linearized modes in extended and critical gravities,''
  Class.\ Quant.\ Grav.\  {\bf 29}, 085017 (2012)
  doi:10.1088/0264-9381/29/8/085017
  [arXiv:1108.5184 [hep-th]].

\bibitem{Lu:2011qx}
  H.~L\"u and K.N.~Shao,
  ``Solutions of free higher spins in AdS,''
  Phys.\ Lett.\ B {\bf 706}, 106 (2011)
  doi:10.1016/j.physletb.2011.10.072
  [arXiv:1110.1138 [hep-th]].

\bibitem{Myers:2010ru}
R.C.~Myers and B.~Robinson,
``Black holes in Quasi-topological gravity,''
JHEP {\bf 1008} (2010) 067
doi:10.1007/JHEP08(2010)067
[arXiv:1003.5357 [gr-qc]].


\bibitem{Liu:2008zf}
J.T.~Liu and W.A.~Sabra,
``Hamilton-Jacobi counterterms for Einstein-Gauss-Bonnet gravity,''
Class.\ Quant.\ Grav.\  {\bf 27} (2010) 175014
doi:10.1088/0264-9381/27/17/175014
[arXiv:0807.1256 [hep-th]].

\bibitem{Fan:2016zfs}
Z.Y.~Fan, B.~Chen and H.~L\"u,
``Criticality in Einstein-Gauss-Bonnet gravity: gravity without graviton,''
Eur.\ Phys.\ J.\ C {\bf 76} (2016) no.10,  542
doi:10.1140/epjc/s10052-016-4389-x
[arXiv:1606.02728 [hep-th]].

\bibitem{Davis:2002gn}
S.C.~Davis,
``Generalized Israel junction conditions for a Gauss-Bonnet brane world,''
Phys.\ Rev.\ D {\bf 67} (2003) 024030
doi:10.1103/PhysRevD.67.024030
[hep-th/0208205].

\bibitem{Liu:2017kml}
H.S.~Liu, H.~L\"u and C.N.~Pope,
``Holographic heat current as Noether current,''
JHEP {\bf 1709} (2017) 146
doi:10.1007/JHEP09(2017)146
[arXiv:1708.02329 [hep-th]].

\bibitem{Li:2018kqp}
  Y.Z.~Li and H.~L\"u,
``$a$-theorem for Horndeski gravity at the critical point,''
Phys.\ Rev.\ D {\bf 97} (2018) no.12,  126008
doi:10.1103/PhysRevD.97.126008
[arXiv:1803.08088 [hep-th]].

\bibitem{Alkac:2018whk}
  G.~Alkac and B.~Tekin,
  ``Holographic $c$-theorem and Born-Infeld gravity theories,''
  arXiv:1805.07963 [hep-th].

\bibitem{Deruelle:2009zk}
  N.~Deruelle, M.~Sasaki, Y.~Sendouda and D.~Yamauchi,
``Hamiltonian formulation of f(Riemann) theories of gravity,''
  Prog.\ Theor.\ Phys.\  {\bf 123}, 169 (2010)
  doi:10.1143/PTP. 123.169
  [arXiv:0908.0679 [hep-th]].

\bibitem{Sisman:2011gz}
  T.C.~Sisman, I.~Gullu and B.~Tekin,
  ``All unitary cubic curvature gravities in $D$ dimensions,''
  Class.\ Quant.\ Grav.\  {\bf 28}, 195004 (2011)
  doi:10.1088/0264-9381/28/19/195004
  [arXiv:1103.2307 [hep-th]].

\bibitem{Oliva:2010eb}
  J.~Oliva and S.~Ray,
  ``A new cubic theory of gravity in five dimensions: Black hole, Birkhoff's theorem and C-function,''
  Class.\ Quant.\ Grav.\  {\bf 27}, 225002 (2010)
  doi:10. 1088/0264-9381/27/22/225002
  [arXiv:1003.4773 [gr-qc]].


\bibitem{Myers:2010jv}
R.C.~Myers, M.F.~Paulos and A.~Sinha,
``Holographic studies of quasi-topological gravity,''
JHEP {\bf 1008} (2010) 035
doi:10.1007/JHEP08(2010)035
[arXiv:1004.2055 [hep-th]].

\bibitem{Bueno:2018xqc}
P.~Bueno, P.A.~Cano and A.~Ruip\'erez,
``Holographic studies of Einsteinian cubic gravity,''
JHEP {\bf 1803} (2018) 150
doi:10.1007/JHEP03(2018)150
[arXiv:1802.00018 [hep-th]].



\end{thebibliography}
\end{document}